\documentclass[%
 amsmath,amssymb,
 aps,
 prb,
 superscriptaddress,
 longbibliography,
 twocolumn,
]{revtex4-2}
\usepackage[T1]{fontenc}
\usepackage{lmodern}
\usepackage{graphicx}% Include figure files
\usepackage{bm}% bold math
\usepackage{xcolor}
\usepackage{xspace}
\usepackage{xfrac}
\usepackage{physics}
\usepackage[expansion=false]{microtype}
\usepackage{tikz}
\usetikzlibrary{arrows.meta, positioning, calc, shadings, decorations.pathmorphing}
\usepackage[pdfusetitle,colorlinks,allcolors=blue]{hyperref}% add hypertext capabilities
\def\equationautorefname#1#2\null{Eq.#1(#2\null)}

\usepackage[T1]{fontenc}
\usepackage[utf8]{inputenc}
\usepackage{amsmath,amssymb,bm,siunitx,graphicx,xcolor,hyperref}
\usepackage[percent]{overpic}
\usepackage{tikz}
\usepackage{physics} 
\usepackage{subfigure} 
\hypersetup{colorlinks=true,linkcolor=blue,citecolor=blue,urlcolor=blue}
  
\begin{document}

%\preprint{APS/123-QED}

\title{Kapitza-like {modulation of near-field radiative heat transfer}}
\author{Mauro Antezza}
\email{mauro.antezza@umontpellier.fr}
\affiliation{Laboratoire Charles Coulomb (L2C), UMR 5221 CNRS-Universit\'{e} de Montpellier, F-34095 Montpellier, France}
\affiliation{Institut Universitaire de France, 1 rue Descartes, Paris Cedex 05 F-75231, France}
\date{\today}

\begin{abstract}
We introduce a Kapitza-like mechanism for the near-field radiative heat transfer and show that fast modulation of any parameter controlling the flux, such as the vacuum gap or a material response, produces a quadratic, time-averaged correction in the slow thermal dynamics. This correction splits into a frequency-independent static term and a low-pass dynamical term, yielding sizable modulation-induced temperature shifts and modified effective thermal conductances that can stabilize or destabilize the steady state. Applying the theory to gap modulation between SiC slabs, we derive analytical scaling laws and predict temperature shifts that are fully measurable with existing experimental platforms, requiring only readily accessible low modulation frequencies of order $\Omega \approx 10^4~\mathrm{rad/s}$. Our results establish a thermal {analog} of the Kapitza mechanism and provide a general route for controlling radiative heat flow in micro- and nanoscale platforms.
\end{abstract}

\maketitle

%%%%%%%%%%%%%%%%%%%%%
\section{Introduction}
%%%%%%%%%%%%%%%%%%%%%
High-frequency parametric excitation can stabilize otherwise unstable equilibria. This phenomenon, known as the Kapitza effect, was analyzed for an inverted pendulum with a rapidly oscillating pivot via a separation of slow and fast motions leading to an effective time-averaged potential~\cite{Kapitza1951,PRX2024}. Averaging and Floquet ideas have since permeated many areas of physics \cite{Bogoliubov1961,Landau1976,Bukov2015,Goldman2014,Eckardt17}.
In parallel, the past two decades have established the foundations and experimental reality of \emph{near-field radiative heat transfer} (NFRHT), where thermally excited evanescent fields enable photon tunneling across subwavelength gaps, yielding heat fluxes far exceeding the blackbody limit~\cite{PolderVanHove1971,Joulain2005,Song2015RHTreview,MessinaAntezzaGenTh,PhysRevB.95.245437,PhysRevApplied.21.024054,Lim2018TailoringNearField,doi:10.1021/acsphotonics.8b01031,lcz1-f5v9}. Quantitative measurements down to nanometric gaps have confirmed the predictions of fluctuational electrodynamics across diverse materials and geometries~\cite{Ottens2011,KimNature2015,Fiorino2018,StGelais2016}. These advances now enable dynamic control via electromechanical gap actuation or dispersion engineering~\cite{Picardi2023Tutorial,YuFan}.
In this work we bring these two threads together and analyze how \emph{fast} periodic modulation of a near-field coupling parameter modifies the \emph{slow} thermal dynamics of a lumped body. In close analogy with Kapitza stabilization, averaging over fast oscillations yields an additional, state-dependent term in the slow equation of motion that can enhance the effective restoring conductance around an equilibrium and considerably affect the thermal state. This opens new modulation mechanisms in currently accessible experimental NFRHT platforms.

\begin{figure}[ht]
    \centering
    % ASSICURATI che qui NON ci sia \resizebox o \scalebox
    \begin{tikzpicture}[
        >=Stealth,
        mainfont/.style={font=\small\bfseries},
        labelfont/.style={font=\footnotesize},
        % Definiamo text width anche negli stili per sicurezza
        plate1/.style={rectangle, draw=blue!80!black, thick, fill=blue!5, minimum width=0.5cm, minimum height=2.8cm, mainfont, align=center},
        plate2/.style={rectangle, draw=orange!80!black, thick, fill=orange!5, minimum width=1.2cm, minimum height=2.8cm, mainfont, align=center}
    ]

        % 1. BODIES
        \node[plate1] (B1) at (-2.4, 0) {\rotatebox{90}{Body 1}};
        \node[plate2] (B2) at (2.4, 0) {Body 2};

        % 2. LABELS & PROPERTIES
        \node[above=2pt of B1, font=\tiny\itshape, blue!80!black] {Lumped body};
        
        % Aggiunta text width esplicita per evitare "stretching"
        \node[left=5pt of B1, labelfont, blue!80!black, align=right, text width=1.2cm] {$T(t)$ \\ $C$};
        
        \node[right=5pt of B2, labelfont, orange!80!black, align=left, text width=1.5cm] {$T_2$ \\ (fixed)};

        % 3. MODULATION
        \node[red!80!black, font=\small\bfseries] (phi) at (0, 2.2) {$\phi(t) = \phi_0 + a\cos(\Omega t)$};
        \draw[->, red, dashed, thick] (phi.south) -- (0, 1.1);

        % 4. RADIATIVE FLUX Q
        \draw[->, blue!80!black, line width=1.8pt, -{Triangle[length=5pt, width=6pt]}] 
            ($(B1.east)+(0,0.8)$) -- ($(B2.west)+(0,0.8)$) 
            node[midway, below=2pt, font=\scriptsize\bfseries] {$Q(T(t),\phi(t))$};

        % 5. BACKGROUND CONDUCTION
        \draw[->, thick, black, decorate, decoration={snake, amplitude=0.6pt, segment length=5pt}] 
            (B1.east) -- (B2.west)
            node[midway, below=3pt, font=\tiny\bfseries] {$G_{\rm cond}(T - T_2)$};

        % 6. GAP DISTANCE d
        \draw[<->, thin] ($(B1.east)-(0,0.9)$) -- ($(B2.west)-(0,0.9)$) 
            node[midway, below, font=\scriptsize] {gap $d$};

        % 7. INPUT POWER
        \draw[<-, thick, green!60!black] ($(B1.west)+(0,0.7)$) -- ++(-0.8, 0) 
            node[left, font=\scriptsize\bfseries] {$P_{\rm in}$};

        % 8. REGIME NOTE
        %\node[font=\tiny, gray] at (0, -1.6) {$G/C \ll \Omega \ll \gamma \ll \omega_{\rm res}$};

    \end{tikzpicture}
    \caption{\label{fig:model} {Schematic of the lumped thermal system. Body~1 [the lumped body with heat capacity $C$ and temperature $T(t)$] and Body~2 (a thicker reservoir at fixed $T_2$) are separated by a gap $d$. The energy exchange is composed of a radiative heat flux $Q$, dynamically modulated by a generic external parameter $\phi(t)$ at frequency $\Omega$, and a background conductive channel $G_{\rm cond}$.}}
    \label{fig:scheme}
\end{figure}
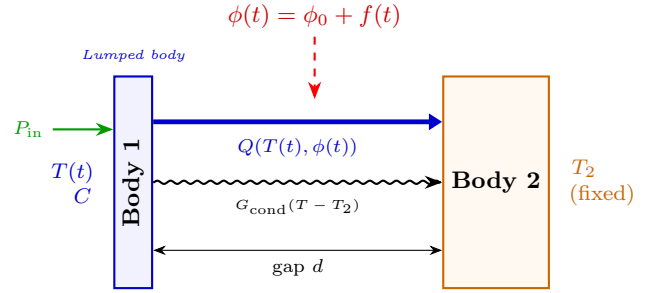

%%%%%%%%%%%%%%%%%%%%%
\section{Physical system}
%%%%%%%%%%%%%%%%%%%%%
We consider two bodies (e.g., parallel plates, sphere-slab,...) where body~1 has a temperature $T(t)$ that evolves under the system's thermal balance, while body~2 is held at a fixed temperature $T_2$, separated by a vacuum gap $d$ (see scheme in Fig.\ref{fig:scheme}). We assume each body remains at a uniform temperature, so body~1 can be treated as a lumped thermal mass with heat capacity per unit area $C$. It receives an external power flux $P_{\rm in}$ and loses heat through a background conductive channel $G_{\rm cond}\,(T - T_2)$, where $G_{\rm cond} > 0$ is the non-radiative thermal conductance per unit area to body~2 (or to the environment). The net radiative flux between the two bodies (W/m$^{2}$) is~\cite{PolderVanHove1971,Joulain2005,Song2015RHTreview,MessinaAntezzaGenTh}
\begin{equation}
 Q(T,\phi) = \int_0^\infty \!\frac{\mathrm d\omega}{2\pi}\, \Phi(\omega,\phi)\,[\Theta(\omega,T)-\Theta(\omega,T_2)],
 \label{eq:Qgen}
\end{equation}
where $\Theta(\omega,T) = {\hbar \omega}/{[\exp(\hbar \omega / k_{\mathrm{B}} T) - 1]}$, $\Phi(\omega,\phi)>0$ is the spectral radiative transmission, and $\phi(t)$ is a generic time-dependent externally-modulated control variable (gap, material property, chemical potential, etc.) affecting the radiative coupling. The dynamics of $T(t)$ is then given by the balance equation:
\begin{equation}
 C\dot T(t) = P_{\rm in} - G_{\rm cond}\;{\big[T(t)-T_2\big]} - Q\big(T(t),\phi(t)\big).
 \label{eq:ode}
\end{equation}
This model applies when the internal thermal resistance is negligible compared to external losses, as in thin membranes or small objects, so that the body equilibrates internally much faster than it exchanges heat with its surroundings. We treat $G_{\mathrm{cond}}$ and $C$ as temperature independent, since the small temperature oscillations considered here render their variations negligible relative to the nonlinear radiative term $Q(T,\phi)$. 
%The modulation $\phi(t)$ is taken to be externally driven and does not inject energy directly; it only modifies the instantaneous radiative flux.

We focus on fast and small modulations around a mean value $\phi_0$:
\begin{equation}
 \phi(t)=\phi_0 + a\cos(\Omega t),\quad 0< a\ll \phi_0,\quad \Omega\gg {|G|/C}.
 \label{eq:mod}
\end{equation}
This modulation induces fast temperature oscillations, motivating the decomposition
\begin{equation}
T(t) = \bar{T}(t) + \xi(t),
\label{eq:sepT}
\end{equation}
where $\bar{T}(t)$ is the \emph{slow} component (varying on the slow timescale $\sim C/{|G|}$), defined as the average of $T(t)$ over one period of the fast drive  $
\bar{T}(t) = (\Omega/2\pi) \int_{t}^{t+2\pi/\Omega} T(t')\, \mathrm{d}t'$,
and $\xi(t)$ captures the small, \emph{fast} oscillatory component  (with a fast timescale $\sim 1/\Omega$, and average  $\langle \xi\rangle=0$ over one modulation period $2\pi/\Omega$).  In Eq.~\eqref{eq:mod} we also introduced the unmodulated effective conductance $G$  evaluated at the average operating point $(\bar T,\phi_0)$: 
\begin{equation}
G(\bar T,\phi_0) = G_{\rm cond} + G_{\rm rad}(\bar T,\phi_0), 
\end{equation}
where $G_{\rm rad}(\bar T,\phi_0)=Q_T(\bar T,\phi_0)\equiv \left.\partial_T Q(T,\phi)\right|_{(\bar{T},\phi_0)}$ is the unmodulated radiative conductance. 

Let us now discuss the timescale requirements for the validity of this model. First, Kapitza-type averaging requires a separation of thermal and modulation timescales: the modulation period must be much shorter than the characteristic thermal timescale {$C/|G|$}, i.e., {$|G|/C \ll \Omega$}. As we show below, this condition does not imply that the magnitude of the modulation-induced effects increases with $\Omega$.

Second, the use in (\ref{eq:ode}) of the static near-field radiative 
heat flux $Q(T,\phi)$ evaluated at the instantaneous value of $\phi(t)$ 
is justified only when the electromagnetic field adjusts 
quasi-instantaneously to the modulation. This requires $\Omega \ll \gamma$, 
where $\gamma$ is the spectral linewidth of the 
dominant evanescent modes, $\omega_{\mathrm{res}}$ being the resonance 
frequency. A separate condition is that 
$\Omega$ must remain well below $\omega_{\mathrm{res}}$, since when 
$\Omega$ approaches $\omega_{\mathrm{res}}$, sideband generation and 
frequency-conversion effects become significant~\cite{YuFan}. Together, 
all these constraints impose the double 
inequality ${|G|}/C \ll \Omega \ll \{\gamma,  \omega_{\mathrm{res}}\}$.
For surface phonon polaritons in SiC, $\omega_{\mathrm{res}} \approx 1.5 \times 
10^{14}$~rad/s and $\gamma \sim 10^{11}$--$10^{12}$~rad/s, both are several orders of magnitude above the modulation frequencies considered here.

%%%%%%%%%%%%%%%%%%%%%
\section{Fast dynamics}
%%%%%%%%%%%%%%%%%%%%%
By inserting $T(t) = \bar T(t) + \xi(t)$ and $\phi(t)$ [Eq.~\eqref{eq:mod}] into the thermal balance Eq.~\eqref{eq:ode} and expanding the radiative flux to second order in $\xi$ and $a$ around $(\bar T, \phi_0)$ gives
\begin{multline}
 C(\dot {\bar T} + \dot \xi) = P_{\rm in} - G_{\rm cond}\;(\bar T+\xi-T_2) \\
 -  \left[ Q_0 + Q_T\,\xi + \tfrac12 Q_{TT}\,\xi^2 + Q_\phi a\cos(\Omega t) \right. \\
 +   \left. Q_{T\phi}\,\xi a\cos(\Omega t)+\tfrac12 Q_{\phi\phi} a^2\cos^2(\Omega t)\right],
 \label{eq:ode22}
\end{multline}
where $Q_0=Q(\bar T,\phi_0)$  and $Q_{\alpha}=\partial_{\alpha} Q$ and $Q_{\alpha\beta}=\partial_{\alpha} \partial_{\beta} Q$ are evaluated at the same point.

{Here, the consistency of the second-order expansion is controlled by the nonlinear scales associated with the derivatives of the heat flux. In particular, the relevant smallness conditions are $|\xi| \ll T_{\rm nl} \equiv \left|Q_T/Q_{TT}\right|$, $|a| \ll \left| Q_\phi/Q_{\phi\phi}\right|$, and $|\xi a| \ll \max\left(|Q_T \xi|,|Q_\phi a|\right)/|Q_{T\phi}|$. Thus, the relevant nonlinear temperature scale is not the absolute temperature, but the scale set by the curvature of the radiative heat flux.}

To perform a two-scale analysis, we first isolate the fast dynamics by retaining terms that either depend on the fast variable $\xi$ or oscillate at the modulation frequency $\Omega$. Terms that vary slowly over a modulation period are effectively constant for the fast dynamics and can be neglected.
Keeping only terms linear in $\xi$ and $a$ yields the equation for the fast component $\xi$:
\begin{equation}
 C\dot\xi + G\xi = -Q_\phi a\cos(\Omega t).
 \label{eqfast}
  \end{equation}
 This is a forced first-order system with sinusoidal forcing. {Under the timescale-separation condition $|G|/C \ll \Omega$, its homogeneous solution evolves on the slow thermal timescale. We therefore focus on the periodic particular solution, which provides the fast periodic response:}
\begin{equation}
 \xi(t) = -\,\frac{Q_\phi\,a}{G^2 + (C\Omega)^2}\,\Big[\,G\cos(\Omega t) + C\Omega\sin(\Omega t)\,\Big].
 \label{eqKapitzasol}
 \end{equation}
The temperature ripple $\xi(t)$ oscillates at the drive frequency $\Omega$, with amplitude filtered by $[G^2+(C\Omega)^2]^{-1}$ due to thermal inertia and conductance. Expressing $\xi(t)$ in polar form gives further insight:
\begin{equation}
\xi(t) = \frac{-Q_\phi a}{\sqrt{G^2 + (C\Omega)^2}}\,
          \cos\!\big(\Omega t - \varphi\big),\;\;
\tan\varphi = \frac{C\Omega}{G}.
\label{eq:xi_phase_form}
\end{equation}
In the frequency domain, writing 
$\cos(\Omega t) = \mathrm{Re}\left[e^{i\Omega t}\right]$ and assuming a harmonic response
$\xi(t) = \xi_\Omega e^{i\Omega t}$, equation~\eqref{eqfast} gives $\xi_\Omega = -H(i\Omega) [Q_\phi a]$,
where $H(i\Omega) = 1/ (G + i C\Omega)$ is a complex transfer function. Thus $1/\sqrt{G^2 + (C\Omega)^2}$  in \eqref{eq:xi_phase_form} is the modulus of $H(i\Omega)$, representing the amplitude response, while the angle $\varphi$ corresponds to the phase lag.  
Physically, $G$ and $C$ play roles analogous to {electrical conductance and electrical capacitance} in an RC circuit, so $\xi(t)$ acts as a thermal low-pass filter.  Slow modulations of $\phi(t)$
($\Omega \ll {|G|}/C$) are transmitted almost in phase  [$\xi(t) \simeq -({Q_\phi a}/{G})\cos(\Omega t)$], so the temperature follows the modulation nearly instantaneously. Fast modulations
($\Omega \gg {|G|}/C$) are strongly attenuated ($\propto 1/\Omega$) and acquire a $\pi/2$ phase shift [$\xi(t) \simeq - ({Q_\phi a}/({C\Omega})) \sin(\Omega t)$] since the finite thermal inertia $C$ delays the response. Hence, the factor $(G^2 + (C\Omega)^2)^{-1}$ in Eq.~\eqref{eqKapitzasol} simultaneously sets the amplitude reduction and the phase lag between forcing and thermal response.

%%%%%%%%%%%%%%%%%%%%%
\section{Averaged slow dynamics}
%%%%%%%%%%%%%%%%%%%%%
Averaging Eq.~(\ref{eq:ode22}) over one modulation cycle using the fast solution (\ref{eqKapitzasol}) yields the slow dynamics of $\bar T(t)$:
\begin{multline}
 C\dot{\bar T} = P_{\rm in} - G_{\rm cond}\,(\bar T-T_2) - Q_0 \\
 - \Delta_{\rm stat}(\bar T) - \Delta_{\rm dyn}(\bar T,\Omega),
 \label{eqKapitzaSolSlow}
 \end{multline}
containing the Kapitza-like correction terms
\begin{eqnarray}
 \Delta_{\rm stat} &=& \frac{a^2}{4} Q_{\phi\phi} ,  \label{eq:Dstat}\\
 \Delta_{\rm dyn} &=& \frac{a^2}{4}\frac{Q_{TT} Q_\phi^2 - 2 Q_{T\phi} Q_\phi G}{G^2 + (C\Omega)^2} ,
  \label{eq:Ddyn}
\end{eqnarray}
with all quantities evaluated at $(\bar T,\phi_0)$. Eq.~(\ref{eqKapitzaSolSlow}) constitutes one of the central results of this work.

The static correction \( \Delta_{\rm stat} \) originates from the flux expansion term
$ \tfrac12 Q_{\phi\phi}a^2\cos^2(\Omega t) $, which, when averaged over a cycle, produces a frequency-independent shift of the mean radiative load, modifying the baseline heat exchange without affecting the frequency response.
The dynamic correction \( \Delta_{\rm dyn} \) arises from the interplay between
the fast temperature ripple \( \xi(t) \) and the modulation of the control parameter.
It involves quadratic averages such as
\(\langle\xi^2\rangle=(Q_\phi a)^2/[2(G^2+(C\Omega)^2)]\) and
\(\langle\xi\cos\Omega t\rangle=-Q_\phi a G/[2(G^2+(C\Omega)^2)]\).
Through the filtering factor \(1/[G^2+(C\Omega)^2]\),
\( \Delta_{\rm dyn} \) depends strongly on the modulation frequency,
acting as a Kapitza-like correction that, as we will see, can either increase or decrease the effective thermal conductance. It vanishes at large $\Omega$, so within the averaging regime  ($\Omega\gg {|G|}/C$) it is largest at the lowest frequency compatible with time-scale separation.

{At this stage, $G(\bar T,\phi_0)$ is not required to be positive: as shown above, the derivation of the fast periodic component and of Eqs.~\eqref{eqKapitzaSolSlow}--\eqref{eq:Ddyn} requires the timescale-separation condition $|G|/C \ll \Omega$ [Eq.~\eqref{eq:mod}], but does not require $G$ to be positive. Hence, Eqs.~\eqref{eqKapitzaSolSlow}--\eqref{eq:Ddyn} remain applicable when $G<0$, as may occur in systems with negative differential thermal conductance~\cite{Sun_2022,doi:10.1021/acs.nanolett.3c03375}, while the stability of the resulting averaged slow dynamics is analyzed in Sec.~\ref{sec:stability}.}

{At this point it is useful to clarify the precise meaning of the Kapitza-like analogy employed throughout this work. In the classical Kapitza pendulum, the effective stabilizing term arises from an inertial second-order equation of motion and increases with the driving frequency. By contrast, the present thermal problem is governed by an intrinsically overdamped first-order relaxation equation, so that the dynamical correction {in Eq.~\eqref{eq:Ddyn}} is filtered by the thermal response function and decreases at sufficiently large frequencies. The analogy with the Kapitza mechanism therefore does not rely on reproducing the specific frequency scaling of the inverted pendulum. Rather, it lies in the common mathematical structure of the averaging procedure: a separation of slow and fast degrees of freedom, the elimination of the fast oscillatory component, and the emergence of an additional quadratic contribution that renormalizes the effective slow dynamics and can modify its stability properties. In this sense, Eqs. \eqref{eqKapitzaSolSlow}-\eqref{eq:Ddyn} represent the thermal counterpart of a Kapitza-type averaged correction, even though its dependence on the modulation frequency differs from that of the mechanical pendulum.
}

%%%%%%%%%%%%%%%%%%%%%
\section{Steady-state mean temperature shift}
%%%%%%%%%%%%%%%%%%%%%
To quantify the effect of the modulation, we compare the steady-state mean temperature $T^*$
of the modulated system, obtained from Eq.~(\ref{eqKapitzaSolSlow}):
\begin{multline}
0 = P_{\rm in} - G_{\rm cond}\,(T^* - T_2) - Q(T^*,\phi_0)\\
    - \Delta_{\rm stat}(T^*) - \Delta_{\rm dyn}(T^*,\Omega),
\label{eq:steady_driven}
\end{multline}
with the steady-state temperature $T_0$ of the unmodulated system:
\begin{equation}
0 = P_{\rm in} - G_{\rm cond}\,(T_0 - T_2) - Q(T_0,\phi_0).
\label{eq:steady_unmod_short}
\end{equation}
Defining the stationary slow-temperature shift $\delta T^* = T^* - T_0$ and 
subtracting Eq.~\eqref{eq:steady_unmod_short} from Eq.~\eqref{eq:steady_driven} one obtains
\begin{multline}
0 = -G_{\rm cond}\,\delta T^*
    - \bigl[\,Q(T^*,\phi_0) - Q(T_0,\phi_0)\bigr]\\
    - \Delta_{\rm stat}(T^*) - \Delta_{\rm dyn}(T^*,\Omega).
\label{eq:steady_diff}
\end{multline}
For weak modulation ($a$ small), expanding the flux to first order in $\delta T^*$ as $Q(T^*,\phi_0)\simeq Q(T_0,\phi_0) + Q_T(T_0,\phi_0)\,\delta T^*$, and keeping
only leading terms in $\delta T^*$ and $a^2$, we obtain the modulation-induced slow-temperature stationary shift
\begin{equation}
\delta T^*
 = -\,\frac{\Delta_{\rm stat}(T_0) + \Delta_{\rm dyn}(T_0,\Omega)}
         {G(T_0,\phi_0)},
\label{eq:deltaT_main}
\end{equation}
where $G(T_0,\phi_0)=G_{\rm cond}+Q_T(T_0,\phi_0)$. Using Eqs.~\eqref{eq:Dstat} {and} \eqref{eq:Ddyn}, this reads explicitly in terms of flux derivatives:
\begin{multline}
\delta T^*
= -\frac{a^2}{4G}
\!\left[
  Q_{\phi\phi}
  + \frac{Q_{TT}\,Q_\phi^2-2\,Q_{T\phi}\,Q_\phi\,G}{G^2+(C\Omega)^2}
\right],
\label{eq:deltaT_explicit}
\end{multline}
with all quantities evaluated at $(T_0,\phi_0)$. Here, $T_0$ itself depends on $\phi_0$ via Eq.\eqref{eq:steady_unmod_short}.
The fast modulation hence produces an $O(a^2)$ shift of the time-averaged steady-state temperature  $\delta T^*$, determined by the nonlinear derivatives of $Q(T,\phi)$ and the frequency-dependent factor contained in $\Delta_{\rm dyn}$.

This result is valid for small modulation amplitudes ($a \ll \phi_0$) and small temperature shifts: $|\delta T^*| \ll |T_0|$ and $|\delta T^*| \ll |T_0 - T_2|$. The latter condition ensures that the linear approximation of $Q(T,\phi)$ around $T_0$ remains valid and that $T^*$ does not approach or cross $T_2$, which would reverse the heat flux and invalidate the linearization. If the shift becomes comparable to the original temperature difference, the fully nonlinear system must be solved.

Regarding the sign of $\delta T^*$, let us consider the case of a vacuum-gap modulation ($\phi=d$). First, $G(T_0,\phi_0)>0$ ensures that $T_0$ is a thermally stable steady state, an assumption required for the linearization leading to Eq.~\eqref{eq:deltaT_main}. 
Concerning the static contribution, in the near field $\Phi(\omega,d)$ typically decreases with $d$ and is convex for small gaps, so that $\Phi_{dd}>0$ and $Q_{dd}>0$, yielding $\Delta_{\rm stat}>0$; the static term  produces a negative mean temperature shift for the considered parameters. To study the dynamic correction, we note that $\Phi_d<0$, and assuming $P_{\rm in}>0$ we also have $T_0 > T_2$, implying $Q_d<0$ and $Q_{Td}<0$. Since $\partial_T^2\Theta>0$, we have $Q_{TT}>0$. 
{Hence, the sign and the magnitude of $\Delta_{\rm dyn}$ are not universal, but are determined by the competition between two contributions with opposite signs, namely: $Q_{TT} Q_d^2 > 0$ and $-2 Q_{Td} Q_d G < 0$. Their relative magnitude depends on the specific thermal configuration, modulation parameters, and on the detailed non-linear dependence of the near-field radiative conductance on temperature and the specific modulation parameter.}

%%%%%%%%%%%%%%%%%%%%%
\section{Stability under modulation}
\label{sec:stability}
%%%%%%%%%%%%%%%%%%%%%
If the averaged slow dynamics admits a stationary solution $T^*$, its stability is determined by considering small perturbations around $T^*$ and linearizing the effective equation. Assuming $|\delta T| \ll T^*$ and  $|\delta T| \ll |T^*-T_2|$, let us replace $\bar T(t)$  with  $T^* + \delta T(t)$ into Eq.~\eqref{eqKapitzaSolSlow}.
Since \(T^*\) is a steady-state solution, it satisfies Eq.~\eqref{eq:steady_driven}.
Subtracting Eq.~\eqref{eq:steady_driven} from Eq.~\eqref{eqKapitzaSolSlow} and
linearizing while retaining only terms linear in \(\delta T\) yields the
perturbation equation:
\begin{equation}
 C\,\delta\dot T
 = -\Big[\,G(T^*,\phi_0)
   + \Delta G_{\rm eff}(T^*,\Omega)\,\Big]\delta T.
 \label{eq:lin_stabilitymain}
\end{equation}
The term in brackets defines the \emph{effective thermal conductance} of the modulated system
\begin{equation}
G_{\rm eff}(T^*,\phi_0,\Omega)
   = G_{\rm cond}
   + Q_T(T^*,\phi_0)
   + \Delta G_{\rm eff}(T^*,\Omega),
   \label{eq:Geff_fin}
\end{equation}
and the (static and dynamic) Kapitza modulation-induced correction to the effective conductance is
\begin{equation}
\Delta G_{\rm eff}(T^*,\Omega)
   =
   \partial_T\!\big[\Delta_{\rm stat}(T)
   + \Delta_{\rm dyn}(T,\Omega)\big]\Big|_{(T^*,\phi_0)}.
\label{eq:def_Geff}
\end{equation}
Equation~\eqref{eq:lin_stabilitymain} governs the evolution of small deviations
\(\delta T(t)\) from the stationary temperature \(T^*\).
The relaxation rate (stability exponent) of the exponential solution $\delta T(t) = \delta T(0) e^{-\lambda t}$ is $\lambda
   = {G_{\rm eff}(T^*,\phi_0,\Omega)}/{C}$. 
Stability requires \(G_{\rm eff} > 0\), ensuring that perturbations decay exponentially.
This behavior is the thermal {analog} of the Kapitza effect in mechanics,
where rapid oscillations generate an averaged stabilizing (or destabilizing)
potential depending on the system parameters.

We note that even when the effective conductance $G_{\rm eff}>0$ stabilizes the averaged slow dynamics while $G<0$, the full nonlinear time-dependent system may still exhibit instabilities beyond the validity of the averaging approximation.

{The SiC example discussed below is not intended to demonstrate such a stability inversion in a specific configuration. Rather, it serves as a realistic platform for illustrating the existence and magnitude of the modulation-induced temperature and conductance corrections. The possibility of {stabilizing an otherwise} unstable averaged thermal equilibrium is expected to be particularly relevant in systems exhibiting negative differential thermal conductance.}

%%%%%%%%%%%%%%%%%%%%%
\section{Gap modulation between parallel plates}
%%%%%%%%%%%%%%%%%%%%%
The main equations above apply to any modulation parameter. As an illustrative case, we consider a gap modulation $\phi = d$. We also use the near-field radiative flux approximation:
\begin{equation}
Q(T,d)= \alpha(d)\bigl(T^m-T_2^m\bigr), \qquad \alpha(d)=\frac{A}{d^n}>0,
\label{eq:simplepowerlaw}
\end{equation}
with typically $m \in [1, 3]$, $n \in [2, 4]$ \cite{Joulain2005,Song2015RHTreview,Pascale, BiehsRMP2021, St-Gelais14, StGelais2016, LUCCHESI} where $n$ and $m$ depend on distances, temperatures, materials, and geometries of the involved bodies.  This power-law form allows closed-form expressions for all quantities of interest, and its three parameters can be determined by fitting the computed or measured flux \(Q(T,d)\).

Using \eqref{eq:simplepowerlaw}, the static correction \eqref{eq:Dstat} becomes 
\begin{equation}
\Delta_{\rm stat}(T)=\frac{n(n+1)}{4}\Big(\frac{a}{d}\Big)^2\,\alpha(d)\bigl(T^m-T_2^m\bigr),
\label{eq:Dstat_general_Power_law}
\end{equation}
which, being  positive, {gives a negative contribution to the stationary temperature shift in Eq.~\eqref{eq:deltaT_main}.}

Using $Q_d=\alpha'S$, $S=T^m-T_2^m$, $Q_{TT}=\alpha m(m-1)T^{m-2}$, and 
$Q_{Td}=mT^{m-1}\alpha'$, Eq.~\eqref{eq:Ddyn} yields
\begin{equation}
\Delta_{\rm dyn}(T,\Omega)
= \frac{n^2(a/d)^2    \alpha(d)^2}{4D(T)}\;F(T),
\label{eq:Ddyn_general_Power_law}
\end{equation}
with $D(T)=G(T)^2+(C\Omega)^2$, $G(T)=G_{\rm cond}+\alpha m T^{m-1}$ and $F(T)=\alpha\,m(m-1)T^{m-2}S(T)^2 - 2mT^{m-1}S(T)\,G(T)$.

Using the model \eqref{eq:simplepowerlaw}, the explicit expression for the thermal conductance shift can be directly derived. The static shift of the effective conductance reads
\begin{equation}
\partial_T\Delta_{\rm stat}\big|_{(T^*,d_0)}
= \frac{n(n+1)}{4}\,\Big(\frac{a}{d_0}\Big)^2\,\alpha(d_0)\,m\,T^{*\,m-1}.
\label{eq:DGstat_general_Power_law}
\end{equation}
This term is independent of the drive frequency $\Omega$ and provides a positive, stabilizing correction to the effective conductance. The dynamic conductance shift is

\begin{multline}
\partial_T\Delta_{\rm dyn}(T,\Omega)\big|_{(T^*,d_0)}
\\
= \frac{n^2}{4}\,\Big(\frac{a}{d_0}\Big)^2\,\alpha(d_0)^2\,
\frac{F'(T^*)D(T^*) - F(T^*)D'(T^*)}{D(T^*)^2},
\label{eq:DGdyn_general_Power_law}
\end{multline}
where $D'(T)=2G(T)G'(T)$,
$G'(T)=\alpha m(m-1)T^{m-2}$,  and  $F'(T) = \alpha m(m-1)\Big[(m-2)T^{m-3}S^2 \Big.
\Big.+ 2T^{m-2}S\,(mT^{m-1})\Big] 
       - 2m\Big[(m-1)T^{m-2}S\,G + 
      T^{m-1}(mT^{m-1})\,G + T^{m-1}S\,G'(T)\Big].$

%%%%%%%%%%%%%%%%%%%%%
\section{Numerical estimates and experimental considerations}
%%%%%%%%%%%%%%%%%%%%%

\begin{figure}[ht!]
\centering
% --- PRIMA RIGA ---
\begin{minipage}{0.23\textwidth}
    \begin{overpic}[width=\textwidth]{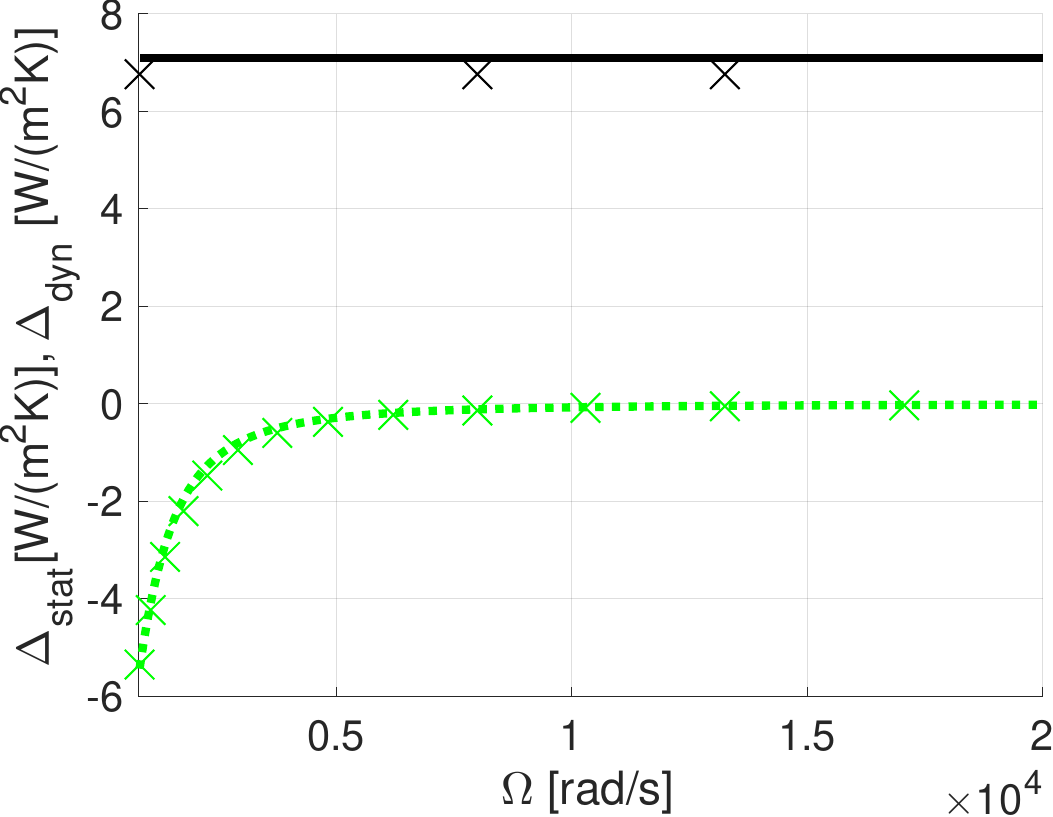}
        \put(85,17){(a)}
    \end{overpic}
\end{minipage}%
\hfill%
\begin{minipage}{0.23\textwidth}
    \begin{overpic}[width=\textwidth]{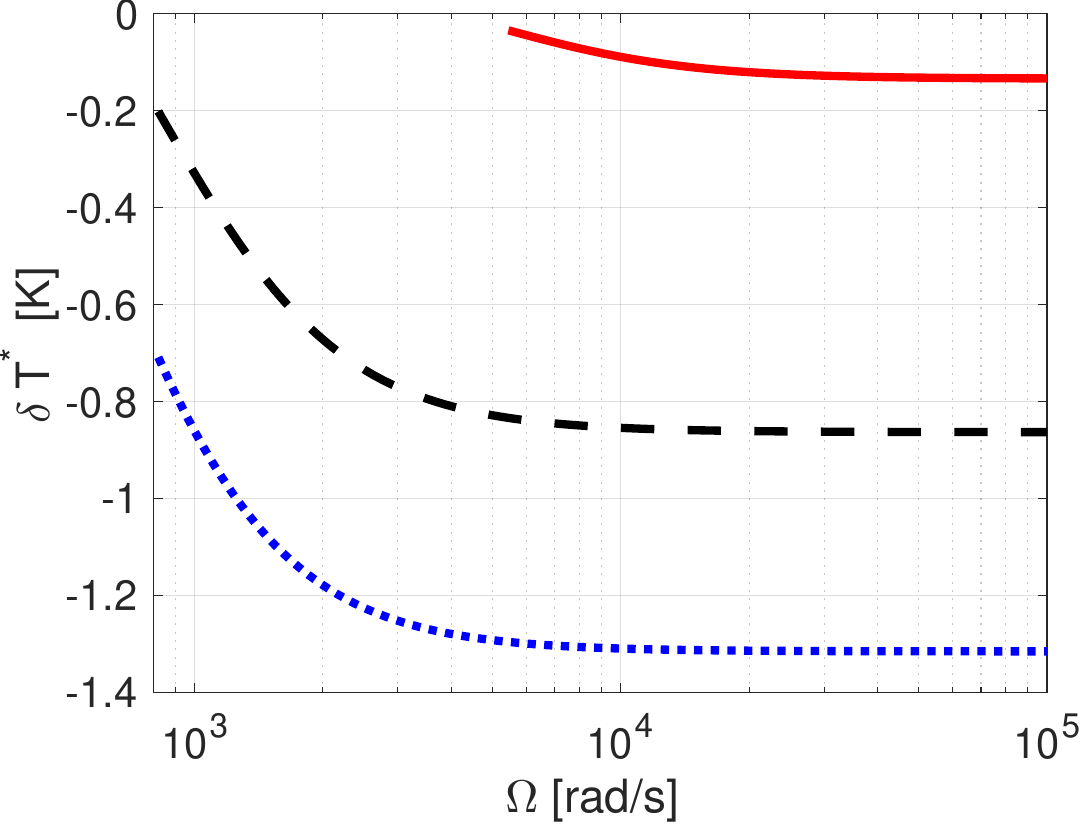}
        \put(15,17){(b)}
    \end{overpic}
\end{minipage}
\vspace{2.4mm}
% --- SECONDA RIGA ---
\begin{minipage}{0.23\textwidth}
    \begin{overpic}[width=\textwidth]{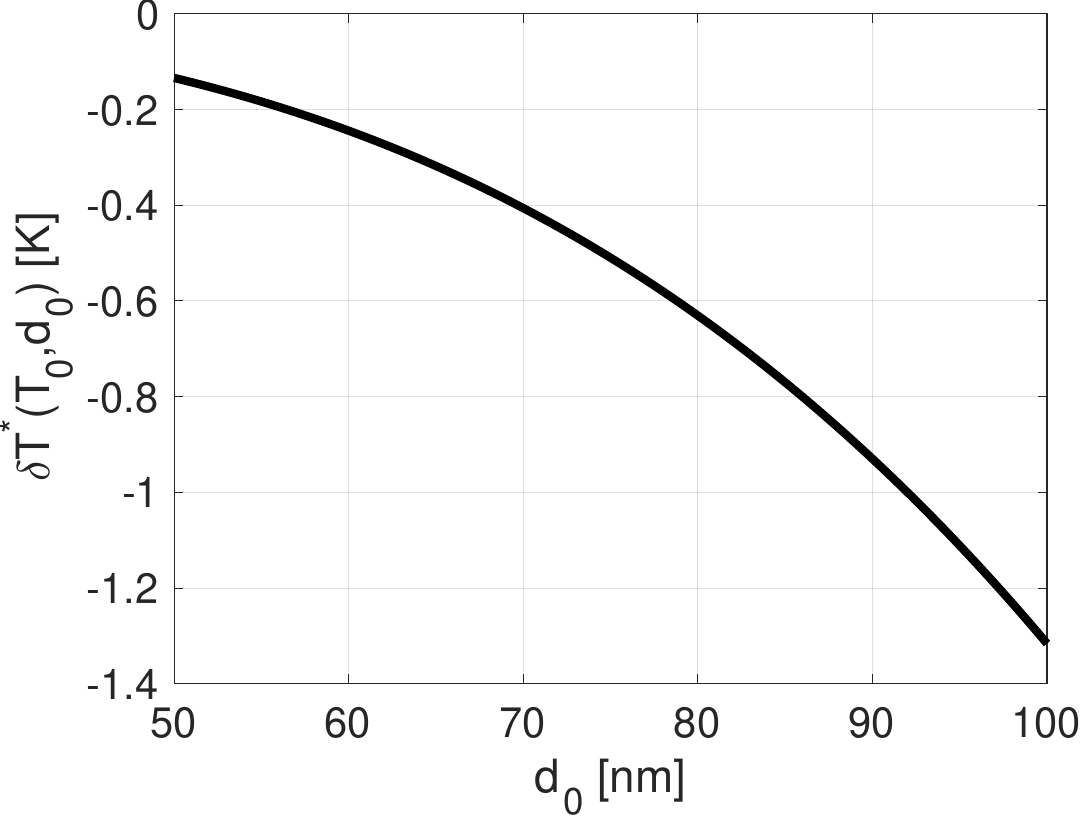}
        \put(85,65){(c)}
        \put(17, 14){\includegraphics[width=0.52\columnwidth]{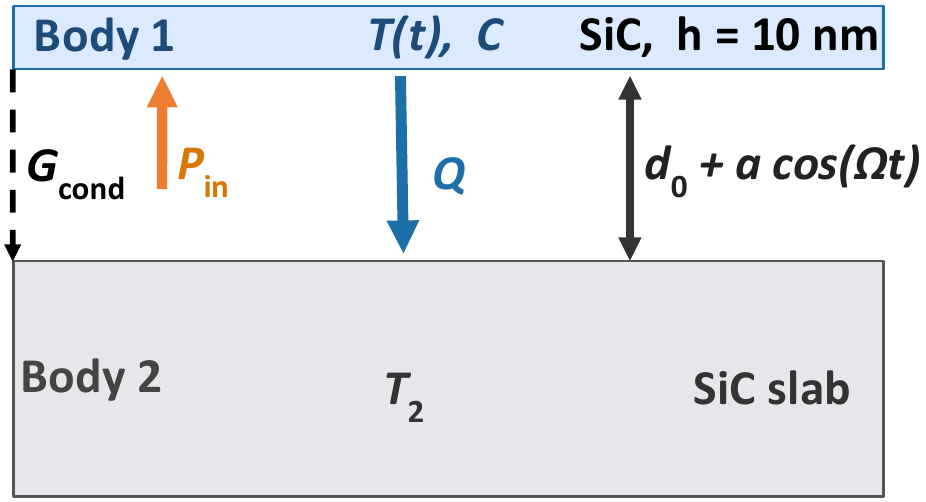}}
    \end{overpic}
\end{minipage}%
\hfill%
\begin{minipage}{0.23\textwidth}
    \begin{overpic}[width=\textwidth]{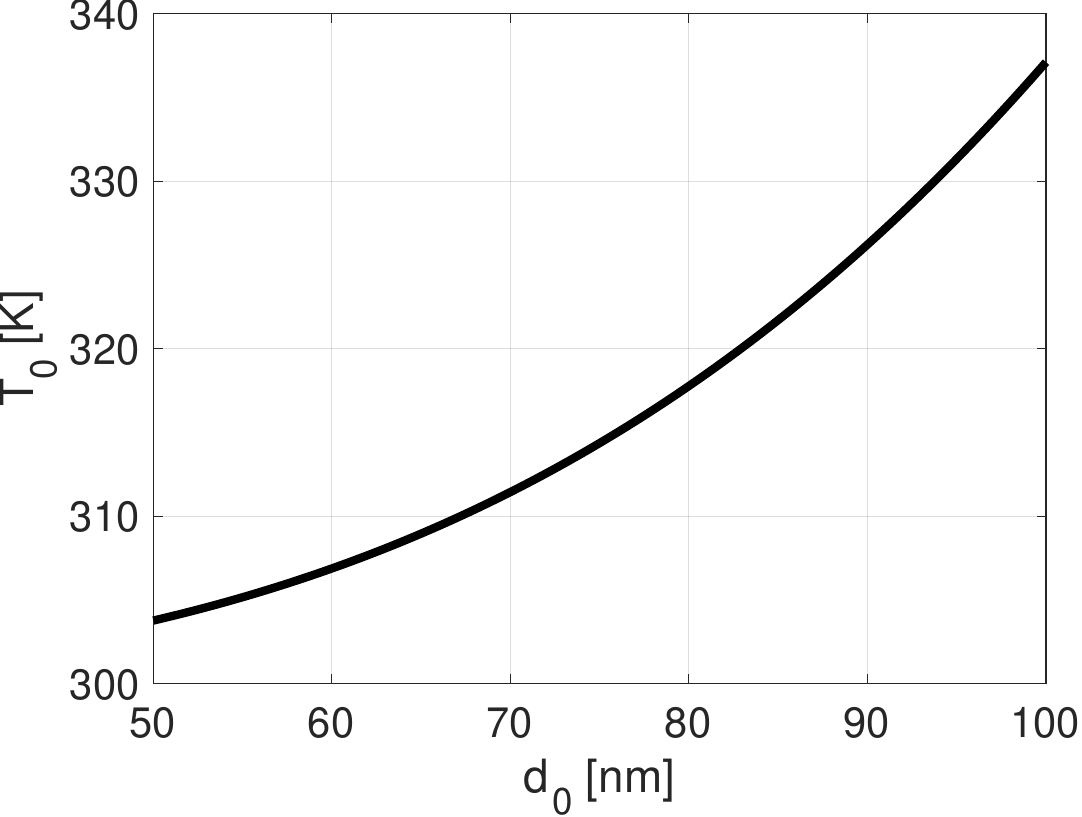}
        \put(15,65){(d)}
    \end{overpic}
\end{minipage}
\caption{
(a) Static (black) and dynamic (green) Kapitza corrections versus $\Omega$ for $d_0=88$~nm.  
(b) Stationary shift $\delta T^*$ versus $\Omega$ for $d_0=50$~nm (red), $88$~nm (black), and $100$~nm (blue).  
(c) High-$\Omega$ limit of $\delta T^*$ versus $d_0$.  
(d) Unmodulated temperature $T_0$ versus $d_0$.  
Parameters: $C=10^{-2}$~J\,m$^{-2}$K$^{-1}$, $G_{\rm cond}=10^{-4}$~Wm$^{-2}$K$^{-1}$, $P_{\rm in}=2\times 10^2$~Wm$^{-2}$, $a=0.1\,d_0$.}
\label{fig:vsd}
\end{figure}

To illustrate these effects, we compute the radiative heat flux $Q(T,T_2,d)$ for two parallel SiC slabs \cite{MessinaAntezzaGenTh,PhysRevB.95.245437}: body~1 is a $h=10$~nm membrane at $T=320$~K, and body~2 has $h_2=100~\mu$m and $T_2=300$~K [see scheme in Fig.~\ref{fig:vsd}(c)]. 
Fitting $Q(T,d)$ near $d_0=88$~nm and $T=320$~K to \eqref{eq:simplepowerlaw} yields $n\approx3.3$, $m\approx1$, and $A\approx4\times 10^{-23}\,\mathrm{Wm}^{\,n-2}\mathrm{K}^{-m}$. 
Using these parameters, we evaluate $\Delta_{\rm stat}(T_0)$ and $\Delta_{\rm dyn}(T_0,\Omega)$ [Eqs.~\eqref{eq:Dstat_general_Power_law} {and} \eqref{eq:Ddyn_general_Power_law}] for $d_0=88$~nm and $a=0.1d_0$ [Fig.~\ref{fig:vsd}(a)], showing the expected low-pass behavior of $\Delta_{\rm dyn}$ for $\Omega\gg G/C\approx820~\mathrm{s}^{-1}$. 
With $C=10^{-2}$~Jm$^{-2}$K$^{-1}$, $G_{\rm cond}=10^{-4}$~Wm$^{-2}$K$^{-1}$, and $P_{\rm in}=2\times 10^2$~Wm$^{-2}$, we obtain $T_0=324$~K and $G(T_0,d_0)=8.22$~Wm$^{-2}$K$^{-1}$. 
The crosses in Fig.~\ref{fig:vsd}(a) are obtained from numerical 
derivatives of the exact flux~\eqref{eq:Qgen}, without relying on the power-law approximation, and indicating a fair agreement with the power-law toy model.

Figure~\ref{fig:vsd}(b) shows the slow-temperature shift $\delta T^*$ [Eq.~\eqref{eq:deltaT_explicit}] versus $\Omega$ for three gaps, $d_0=50,\,88,\,100$~nm (with corresponding $T_0=303,\,324,\,337$~K) and $a=0.1d_0$. As expected, $\delta T^*$ is always negative and saturates at large $\Omega$, giving $\delta T^*=-0.13$~K, $-0.86$~K, and $-1.32$~K, respectively.  
These shifts can be strongly amplified by increasing the input power.  
For instance, keeping the same power-law parameters for $Q$ and $d_0=88$~nm, raising the input to $P_{\rm in}=2\times 10^3\,\mathrm{Wm^{-2}}$ increases the unmodulated temperature to $T_0=542$~K and yields a much larger Kapitza-like shift, $\delta T^*=-8.6$~K.

Figure~\ref{fig:vsd}(c) shows the temperature shift $\delta T^*$ versus $d_0$ in the large-$\Omega$ limit, where $\Delta_{\rm stat}(T_0)$ dominates. The shift decreases at short distances because, in Eq.~\eqref{eq:deltaT_explicit}, the denominator (temperature derivative) decreases more slowly than the numerator (distance derivative), given the fitted near-field power-law exponents $n$ and $m$. In the far field ($n=0$, $m=4$), both $\Delta_{\rm stat}$ and $\Delta_{\rm dyn}$ vanish, so $\delta T^*=0$.  
Figure~\ref{fig:vsd}(d) reports the unmodulated temperature $T_0$ versus $d_0$.

\begin{figure}[ht!]
\begin{overpic}[scale=0.20]{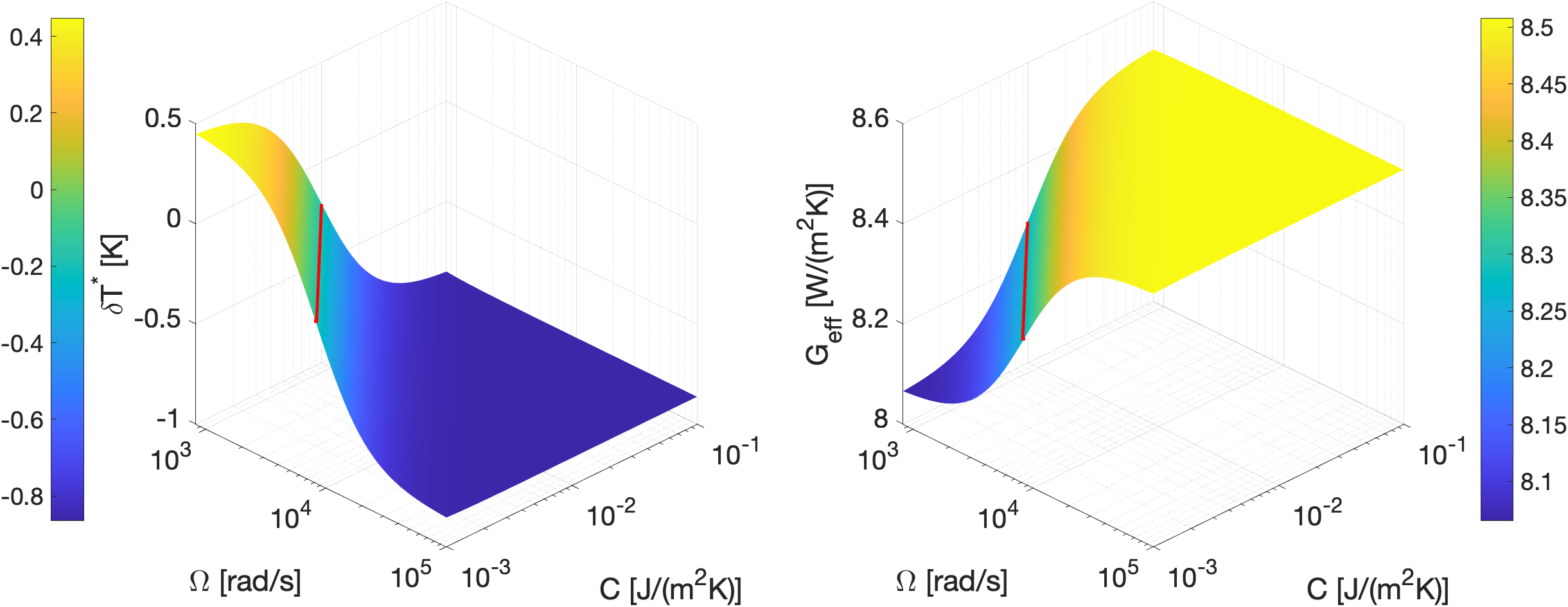}
    % Posizioni in percentuale [x, y]
    \put(10,37){(a)}  % lettera per il primo subplot
    \put(55,37){(b)}% lettera per il secondo subplot
\end{overpic}
\caption{
(a) Slow-temperature shift $\delta T^*$ and (b) effective conductance $G_{\rm eff}(T^*,d_0,\Omega)$ versus $\Omega$ and $C$, for $d_0=88$~nm ($T_0=324$~K), using the same parameters as in Fig.~\ref{fig:vsd}.  
Red curves indicate the limit of validity of the averaging approximation $\Omega = G(T_0,d_0)/C$, with $G(T_0,d_0)=8.22$~Wm$^{-2}$K$^{-1}$.}
\label{fig:3D}
\end{figure}
\begin{figure}[ht!]
\begin{overpic}[scale=0.45]{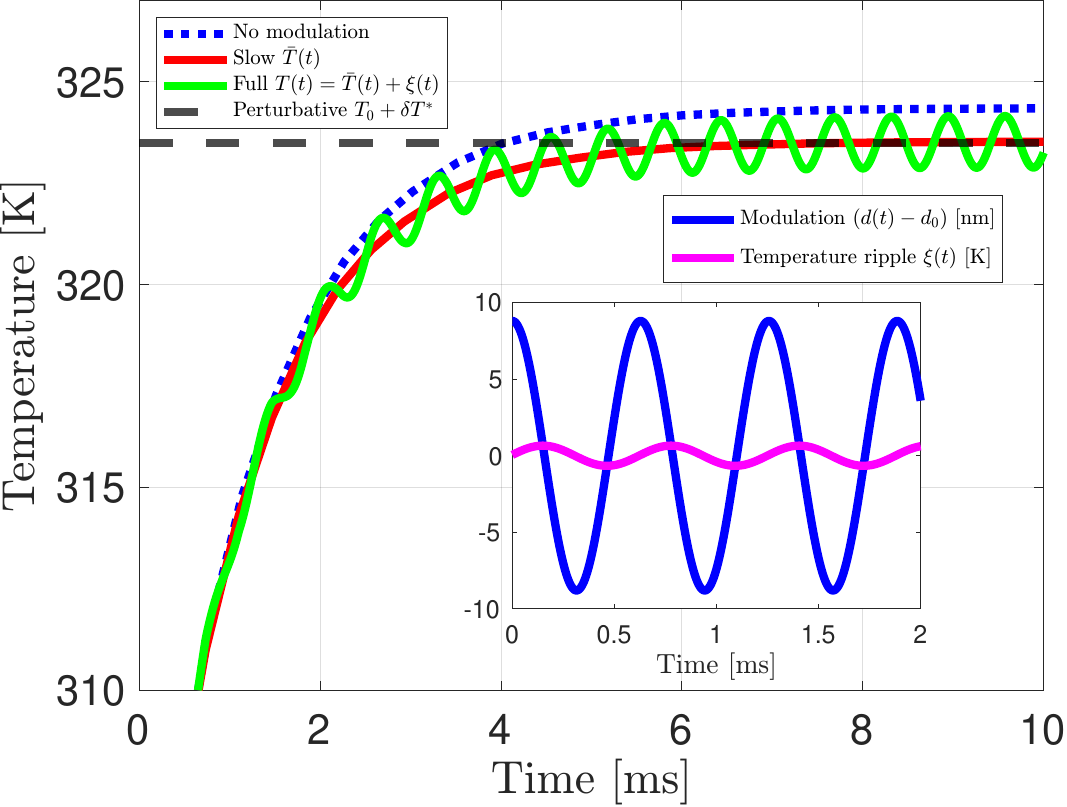}
\end{overpic}
\caption{Time evolution for 
$d_0 = 88$~nm, $\Omega = 10^4$~rad/s, and the same 
parameters as in Fig.\ref{fig:vsd}: unmodulated steady 
state $T_0$ [(\ref{eq:steady_unmod_short})]; slow averaged component $\bar{T}(t)$ 
[(\ref{eqKapitzaSolSlow})]; full solution 
$T(t) = \bar{T}(t) + \xi(t)$ [(\ref{eqKapitzasol}) and (\ref{eqKapitzaSolSlow})]; perturbative 
prediction $T_0 + \delta T^*$ [\eqref{eq:deltaT_main}]
Inset: fast oscillations of the gap modulation 
$d(t) - d_0$ and the induced {dephased} temperature 
ripple $\xi(t)$.}
\label{fig:time}
\end{figure}

{It is worth noting that, for the parameters considered here, we find $T_{\rm nl}=\left| Q_T/Q_{TT}\right|\simeq 148\,{\rm K}$, which is much larger than \(\xi_{\max}=0.66\,{\rm K}\). The nonlinear scale associated with the modulated parameter is $\left| Q_d/Q_{dd}\right|\simeq 21.4\,{\rm nm}$, 
which is larger than the modulation amplitude \(a=8.8\,{\rm nm}\). Finally, the mixed term satisfies $|Q_{Td}\,\xi_{\max} a| /
\max\left(|Q_T\xi_{\max}|,|Q_d a|\right) \simeq 3.0\times 10^{-2}$. These estimates show that the consistency conditions of the second-order expansion are satisfied in the parameter regime considered here, although the pure nonlinear correction in the modulated parameter represents the largest second-order contribution, as expected for a near-field heat flux with a strong distance dependence.}

Figures~\ref{fig:3D}(a) and \ref{fig:3D}(b) display the slow-temperature shift $\delta T^*$ and the effective conductance $G_{\rm eff}(T^*,d_0,\Omega)$ [Eq.~\eqref{eq:Geff_fin}] as functions of the modulation frequency $\Omega$ and the heat capacity $C$, for $d_0=88$~nm (all parameters given in the caption). The red curve indicates $\Omega = G(T_0,d_0)/C$, marking the lower bound of the high-frequency averaging regime ($\Omega \gg G(T_0,d_0)/C$). Reducing $C$ enhances the Kapitza-like effect. 

{Finally, Fig.~\ref{fig:time} presents the full temporal dynamics of the system, with further details provided in the caption. We see that the perturbative prediction for $\delta T^*$ agrees with the long-time stationary value obtained from the time-dependent solution of the averaged equation.}

We note that the lumped-body approximation is well justified for our 
$h = 10$\,nm SiC membrane: the Biot number $\mathrm{Bi} = Gh/k 
\approx 7\times 10^{-10} \ll 1$ (where $k \approx 120$--$490$\,W\,m$^{-1}$K$^{-1}$ 
is the SiC thermal conductivity) confirms static thermal uniformity, 
and the thermal penetration depth $\delta_{\mathrm{th}} = 
\sqrt{2\kappa/\Omega} \approx 32\,\mu$m at $\Omega = 10^5$\,rad/s 
[where $\kappa = k/(\rho c_p) \approx 5\times 10^{-5}$\,m$^2$s$^{-1}$ 
is the thermal diffusivity, with $\rho \approx 3210$\,kg\,m$^{-3}$ 
and $c_p \approx 750$\,J\,kg$^{-1}$K$^{-1}$] exceeds $h$ by more 
than 3 orders of magnitude, ensuring a uniform dynamic response 
throughout the entire modulation frequency range considered. {{Larger thicknesses} or larger frequencies are still acceptable provided $h\sqrt{\Omega} \ll 10^{-2}$ms$^{-1/2}$.
We also stress that all the asymptotic high-frequency {limits} discussed in the present model should be interpreted as a formal limit of the reduced-order lumped-body description rather than as a physically attainable regime. It does not imply dissipation-free operation at arbitrarily high mechanical driving frequencies, since at sufficiently large modulation frequencies, spatial thermal gradients, thermoelastic effects, and additional dissipative mechanisms may become relevant, requiring a fully distributed {thermomechanical} treatment beyond the scope of the present work.}

Promising platforms include polar dielectrics supporting surface phonon polaritons~\cite{Joulain2005,BiehsRMP2021}, ultrathin suspended membranes minimizing $C$, and {actuators based on microelectromechanical systems (MEMS) or nanoelectromechanical systems (NEMS)}, enabling stable sub-100\,nm gaps~\cite{StGelais2016}. The predicted shifts $\delta T^*$ and changes in $G_{\rm eff}$ are directly measurable in suspended-membrane NFRHT experiments. 

{In practical implementations, actuator-related dissipation and residual parasitic heating may contribute additional terms to the thermal balance \eqref{eq:ode}. For typical capacitive MEMS/NEMS actuation schemes driven at kHz-MHz, the electrical dissipation can remain very small (typically in the {nanowatt-microwatt} range under vacuum operating conditions) due to the predominantly capacitive nature of the driving mechanism, yielding a total power much smaller than the radiative flux over the same area, and therefore does not qualitatively affect the thermal balance discussed here.}

Temperatures can be read out via resistance or optical thermometry, both offering $10$-$50$\,mK resolution, well below the predicted $\sim1$\,K shifts. 
A practical protocol is to sweep the modulation frequency $\Omega$ while monitoring the slow temperature component $\bar T(t)$; the crossover at $\Omega \simeq G/C$ and the high-$\Omega$ saturation of $\delta T^*$ provide clear signatures of the Kapitza-like averaging mechanism. These considerations indicate that the effect is well within reach of current experimental platforms.

{It is worth stressing that, in a mechanically actuated implementation, additional conservative force fields, such as the Casimir interaction, may induce static deflections or distort the prescribed harmonic trajectory. In the present reduced thermal model, however, d(t) is treated as an externally imposed modulation; a realistic device design would require solving the coupled electromechanical problem, while the resulting waveform could be directly incorporated into the same radiative heat-transfer framework.
The parallel-plate modulation considered here should be regarded as a minimal proof-of-concept geometry aimed at isolating the Kapitza-like radiative mechanism. Other experimentally relevant configurations, such as sphere-plane geometries, may be more suitable for practical mechanical implementations by reducing constraints associated with parallelism, deformation, and alignment. It is worth emphasizing that the proposed Kapitza-like effect is not restricted to mechanical gap modulation. Since the mechanism only requires a time-dependent modulation of the radiative heat flux, it could also be implemented by modulating the dielectric response of one of the bodies, which may alleviate some practical issues related to mechanical actuation, alignment, and deformation.}\\

%%%%%%%%%%%%%%%%%%%%%
\section{Conclusions}
%%%%%%%%%%%%%%%%%%%%%
We have generalized the Kapitza mechanism to NFRHT, showing that rapid modulation of any flux-controlling parameter produces a universal correction to the slow thermal dynamics. The resulting closed-form dynamics reveal two contributions: a frequency-independent static term and a frequency-dependent dynamical term, which together generate a modulation-induced temperature shift and modify the effective thermal conductance. As in the Kapitza pendulum, fast driving produces an effective potential; here it yields an \emph{effective thermal conductance} that can stabilize or destabilize the temperature evolution. This mechanism can be particularly useful when the unmodulated system is only marginally thermally stable. The theory is fully general and applies to arbitrary geometries and to modulation of dielectric properties or separations, requiring only small amplitudes and a clear timescale separation. For gap modulation between two parallel slabs, we derived analytical scalings and numerically predicted temperature shifts of $\approx 1$ K for ultrathin SiC membranes, with $\approx 10$ K shifts readily accessible at higher powers. We also identify a promising experimental platform for observing this effect and show that it lies within current experimental feasibility, requiring accessible modulation frequencies of order $\Omega \approx 10^4~\mathrm{rad/s}$.  Future investigations could explore promising platforms for implementing effective thermal-conductance stabilization in near-field systems exhibiting negative differential thermal conductance, enabled, for instance, by the metal--insulator transition of phase-change materials such as vanadium dioxide (VO$_2$). {These results establish a Kapitza-like averaging mechanism for near-field radiative heat transfer, whereby fast modulation generates additional effective terms in the slow thermal dynamics through the elimination of fast oscillatory degrees of freedom.}\\

\bibliography{kapitza_nf}

\end{document}